\documentclass[12pt]{article}
\usepackage{float}
\usepackage{amsmath}
\usepackage{amsfonts}
\usepackage{amssymb}
\usepackage{graphicx}
\usepackage{cite}
\usepackage{color}
\usepackage{dcolumn}
\usepackage{wrapfig}
\usepackage{longtable}
\usepackage{times}
\usepackage{caption}
\usepackage{subcaption}
\usepackage{bm}
\usepackage[margin=0.7in]{geometry}
\usepackage{newfloat}
\DeclareFloatingEnvironment[name={SM Figure}]{suppfigure}
\DeclareFloatingEnvironment[name={SM Table}]{supptable}
\begin{document}
\begin{flushleft}
{\LARGE
\textbf{Emergence of chimera in multiplex network}
}
\\
\vspace{0.3cm}
\bf Saptarshi Ghosh$^{1\dagger}$,
Sarika Jalan$^{1,2,\ast}$,
\\
\vspace{0.2cm}
\it ${^1}$ Complex Systems Lab, Discipline of Physics, Indian Institute of Technology Indore, Khandwa Road, Simrol,
Indore 452020, India
\\
\it ${^2}$ Centre for Biosciences and Biomedical Engineering, Indian Institute of Technology Indore,  Khandwa Road, Simrol,
Indore 452020, India
\\
${^{\dagger}}$ Email: sapta15@gmail.com
\\
${^{\ast}}$ Email: sarika@iiti.ac.in (Corresponding author)
\end{flushleft}
\begin{abstract}
Chimera is a relatively new emerging phenomenon where coexistence of 
synchronous and asynchronous state is observed in symmetrically coupled dynamical units. We report observation of the chimera state in multiplex networks where individual layer is represented by 1-d lattice with non-local interactions. 
While, multiplexing does not change the type of the chimera state and 
retains the multi-chimera state displayed by the isolated networks, it changes the regions of the incoherence. We investigate emergence of coherent-incoherent 
bifurcation upon varying the control parameters, namely, the coupling strength and the network size. Additionally, we investigate the effect of initial 
condition on the dynamics of the chimera state. Using a measure based on
the differences between the neighboring nodes which distinguishes 
smooth and non-smooth spatial profile, we find
the critical coupling strength for the transition to the chimera state. Observing chimera in a multiplex network with one to one inter layer coupling is important to gain insight to many real world complex systems which inherently posses multilayer architecture.
\end{abstract}


\section{Introduction}
In past few decades, network science has discovered a plethora of novel phenomena while trying to mimic real world systems in a better manner. One of such discovery is an observation of the chimera state. It was first reported by Kuramato et. al. in 2002 while investigating non locally coupled identical oscillators in a ring network ~\cite{kuramoto_2002}. Later, it was analyzed and christened by Abrams and Strogatz in 2006 as chimera state ~\cite{Abrm_strgz_2006}. Like, its counterpart in Greek mythology, a chimera state has come to be referred as a mathematical hybrid state in which coherent and incoherent dynamics coexist in non-locally coupled identical oscillators in a structurally symmetric network.

Chimera has been extensively investigated both theoretically 
\cite{theo1,theo2,theo9} and experimentally \cite{exp1,exp2}.
It has been observed in plenty of networks including phase oscillators \cite{phase1,Abrm_strgz_2006,phase2,phase3,theo1,theo2}, 
chemical \cite{chemosc_1,chemosc_2}, mechanical oscillators \cite{mechosc}, 
neuron models \cite{neuronosc}, planar oscillators \cite{planarosc}, 
boolean networks \cite{booleanosc_1,booleanosc_2}, 1D superconducting meta 
material \cite{supcondcosc}, etc. Chimera was originally reported for 
non-locally coupled oscillators, recently it has been reported in feed back delayed networks \cite{fdbck_1,fdbck_2}, globally coupled networks 
\cite{gcm_1,gcm_2}, time varying networks \cite{tmvarosc} and networks with 
purely local coupling \cite{local_coup}. Moreover, different types of chimera 
has been reported including multi cluster chimera 
\cite{mult_clus_1,mul_clus_2}, virtual \cite{exp2}, breathing \cite{breathing_c} and two dimensional chimera \cite{phase3,2D_1}.  A recent work suggests 
emergence of Chimera, dependent on nonhyperbolicity of dynamical systems for 
both the time-discrete and time continuous cases \cite{hyperbolocity}.

The chimera state has also been reported for various real world networks models such as Rosenzweig-MacArthur oscillators for ecological networks \cite{ecological_network}.
Chimera has also been characterized by the state of the dynamical evolution of 
the network. Type - I chimera is characteristic of the hyper chaotic behavior 
with many positive Lyapunov exponents \cite{spectral}. This type of chimera has primarily been observed for time-continuous systems like complex Ginzburg-Landau oscillators or Kuramoto oscillators.  
In Type-II chimera, only spatial chaos has been observed with a rather simple temporal behavior (mostly periodic). Though, this type of chimera has been 
reported mainly for the time-discrete systems (maps) 
\cite{theo9,tmdscrtmp,exp1}, it has recently been observed for the 
time-continuous Stuart-Landau oscillators as well \cite{slosc}.

\begin{figure}[t]
\centerline{\includegraphics[width=1.0in, height=1.0in]{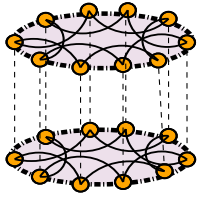}}
\caption{Schematic diagram for multiplex network consisting of two layers. 
Each layer is represented by 1D lattice with non-local interaction. 
Each node (circle) has the same coupling architecture.}
\label{fig.mul_pic}
\end{figure}

Further, modeling real world complex systems under the multiplex network 
framework is one of the recent advancements in the network theory 
\cite{mul_review,mul_review_2,mul_review_3,mul_review_4}. We consider a multiplex network consisting of the same nodes across the layers (Fig.~\ref{fig.mul_pic}) and 
investigate the occurrence of chimera state in the multiplex ring networks. 
In this structure, each node has exactly the same connection architecture. 
We observe Type-II chimera with spatial chaos and periodic temporal behavior. 
Though, the chimera state, upon multiplexing, remains of the same type as observed for the isolated network, the multiplexing changes the region of 
incoherence. Dependence of the chimera state on initial conditions is observed 
for the multiplex networks as already observed for the isolated networks. 
Additionally we present a measure in terms of distance variable to distinguish between the coherent and the chimera state and to identify the transition point 
for the coherent-incoherent bifurcation. We also investigate role of
the size of the network in determining the critical coupling strength for the symmetry breaking and thus emergence of the chimera state. 

\section{Model}
We consider a multiplex $S^1$ ring network with N nodes in each layer, where $S^1$ represents one dimensional symmetric cyclic group with elements being invariant to the permutation operation \cite{S1_grp}. Considering $z_t(i),i=1,...,mN$ as a real dynamical variable at time $t$ for the $i^{th}$ node, the dynamics of the network can be described as,
\begin{equation}
z_{t+1}(i)=f(z_t(i))+\frac{\varepsilon}{(2rN+1)} \sum_{j=1}^{mN} A_{ij} [ f(z_t(j))-f(z_t(i)) ]
\label{eq.evol}
\end{equation}
where $\varepsilon$ represents the coupling strength, $m$ represent number of layers, $r$ represents the coupling radius defined by $r=P/N$, with $P$ signifying the number of neighbours in each direction in a layer. The elements of the adjacency matrix A of a network is defined as $A_{ij}=1$ or $0$ depending upon whether $i^{th}$ and $j^{th}$ nodes are connected or not. The diagonal entries $A_{ij}=0$ represents no self connection in the network. The adjacency matrix $A$ for the multiplex network can be written as, 
$$
   A=
      \begin{pmatrix} 
      A^1 & I & .   & . & . & I  \\ 
      I & A^2 & I   & . & . & I  \\
      . &   I & A^3 & . & . & .  \\
      . &   . &  .  & . & . & .  \\
      . &   . &  .  & . & . & .  \\
      I &   . &  .  &  .& . & A^m  \\
      \end{pmatrix}, 
$$
where $A^1$,$A^2$,...,$A^m$ represent the adjacency matrix of the first, second,\ldots,$m^{th}$ layer, respectively and
$I$ is an unit NxN matrix. 
Note that the number of nodes in each layer of the multiplex network is same. A mismatch in the network size of the layers will yield nodes in different layers having different interaction patterns and hence we can not define the chimera state. We use logistic map $f(z)=\mu z(z-1)$ with the 
bifurcation parameter $\mu =3.8$ at which individual logistic map
exhibits chaotic behavior. We consider coupling radius $r = 0.32$ for each 
layer indicating degree of each node being 64.
A state of the network is defined as spatially coherent if for any node $i \in S^1$, the spatial distance between the neighboring nodes approaches to zero for $t \rightarrow \infty$
\begin{equation}
\lim\limits_{t \rightarrow \infty} \mid{z_t(i+1) - z_t(i)}\mid \rightarrow 0, \, \, \, \, \forall \; i \in S^1
\label{eq.cohr}
\end{equation}
Geometrically, this signifies a smooth profile of the spatial curve. Smoothness of the curve, signifying the correlated spatial values of the neighboring nodes, is defined with the absence of any discontinuity in the spatial curve. Whereas, temporal coherence (synchronization) is defined as, 
$
\lim\limits_{t \rightarrow \infty} \mid{z_{t}(j) - z_t(i)}\mid \rightarrow 0$ for $\forall \; i,j \in S^1$.
Therefore, temporal coherence can be written as $z_t(1)=z_t(2)=......= z_t$ which leads to a straight line for the spatial curve associated with the temporal coherence. 
\begin{figure}[t]
\centering
\includegraphics[width=3.4in, height=3.5in]{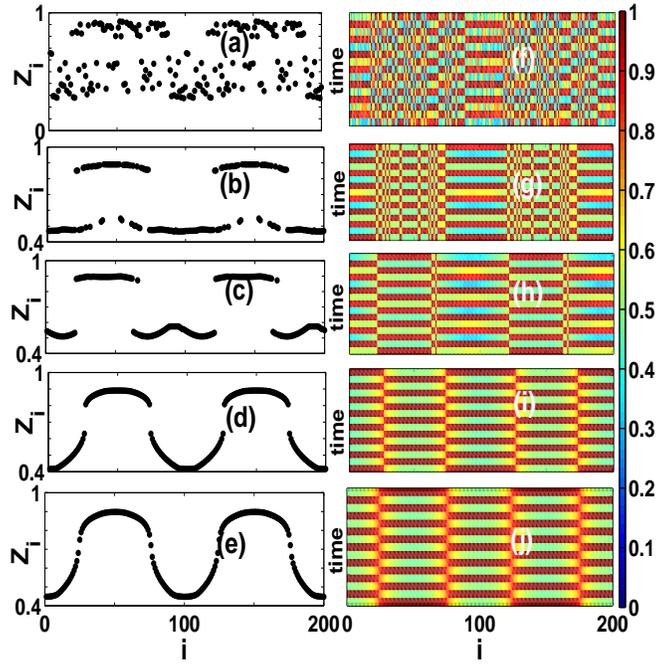}
\caption{Snapshots and spatio-temporal plots for multiplex $S^1$ ring network.  (a) and (f) are for $\varepsilon=0.1$, (b) and (g) $\varepsilon=0.28$, (c) and (h) 
$\varepsilon=0.30$, (d) and (i) $\varepsilon=0.40$, (e) and (j) 
$\varepsilon=0.44$. Number of nodes in each layer remains $N=100$ and
coupling radius $r=0.32$. Initial transient is taken as 5000. results are presented for time range 5000 to 5015.}
\label{fig.main}
\end{figure}
Appearance of discontinuity in the smooth spatial curve implies coexistence of 
the coherence and incoherence regions. To demonstrate the absence of smoothness, 
we define a measure based on the
spatial distance as follows,
\begin{equation}
d(i)=(z(i+1)-z(i))-(z(i)-z(i-1))
\label{eq.dmsr}
\end{equation}
which captures the difference of the spatial distances between the neighboring 
nodes. For smooth spatial profile $d(i)=0$, signifying a 
symmetric distribution of the distances between neighboring nodes, while discontinuity in the spatial profile, signifying the transition point, is indicated 
as kinks in the distribution. We use this measure to find the critical coupling strength for the symmetry breaking and thus resulting the chimera state.

\section{Coherent-Incoherent bifurcation}
\begin{figure}[t]
\centering
\includegraphics[width=3.5in, height=2.0in]{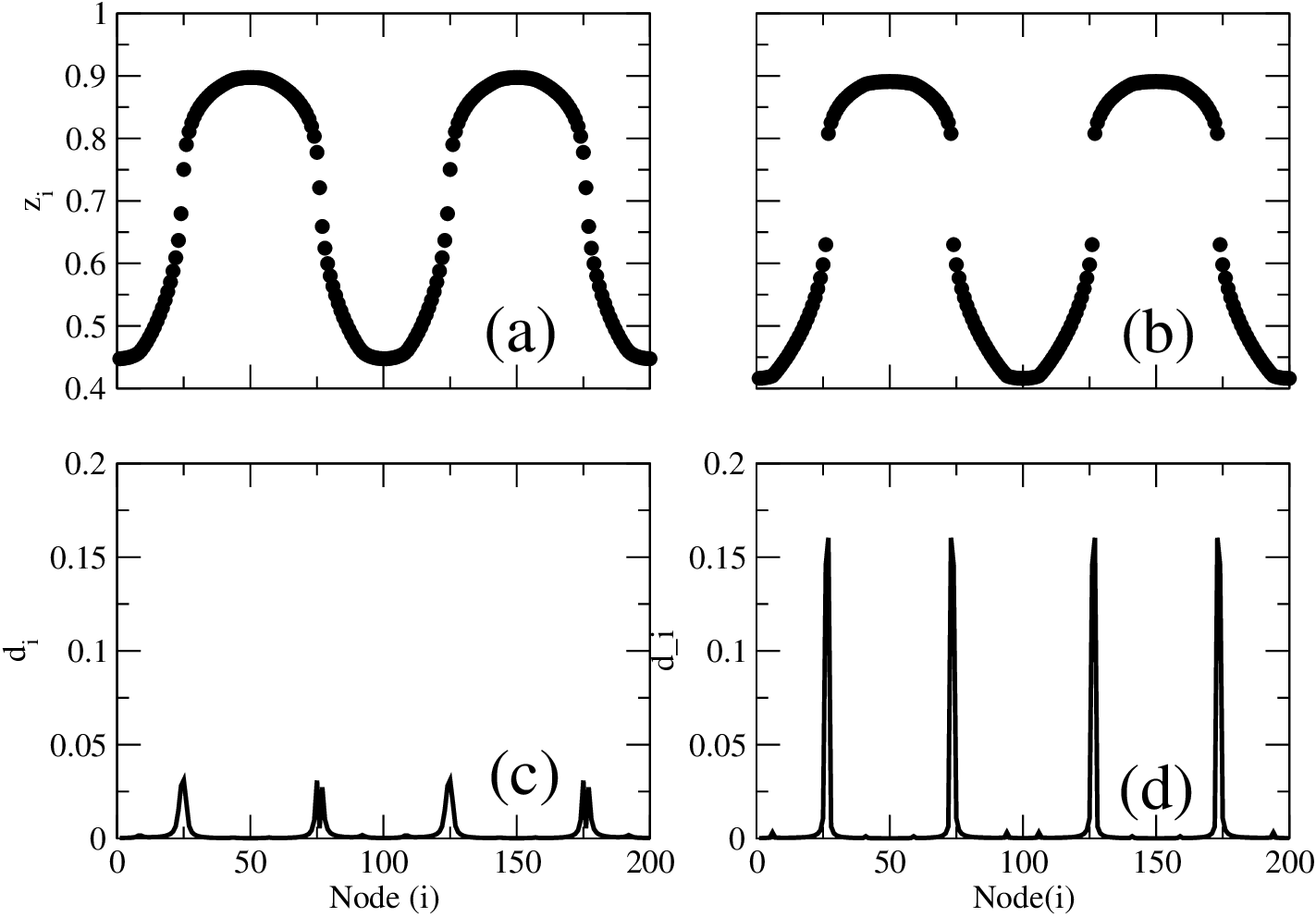}
\caption{Distance measure for multiplex $S^1$ ring network (a),(c) for
$\varepsilon=0.44$ and (b),(d) for $\varepsilon=0.4$. Network size is $N=100$ 
in each layer and coupling range is $r=0.32$.}
\label{fig.dismsr1}
\end{figure}

We evolve Eq.~\ref{eq.evol} starting with a set of special 
initial conditions and after an 
initial transient, study the spatio temporal patterns of the multiplex network. 
Note that uniform or the Gaussian distributed random initial conditions 
lead to either a completely coherent, spanning all the nodes, or a completely 
incoherent
state depending upon the coupling strength. For $0.28 \le \varepsilon \le 0.32$
leads to the incoherent evolution of all the nodes. Motivated from
\cite{kuramoto_2002,Abrm_strgz_2006}, we use a hump back function to generate
initial conditions as follows.
We choose an uniform random number 
$z_0(i)$ for initial state for $i^{th}$ node
within some interval which varies like a Gaussian function as:
\begin{equation}
z_0(i)=exp\big[-\frac{(i-\frac{N}{2})^2}{2\sigma^2}\big]
\label{ini_eq}
\end{equation}
The variance $\sigma$ is chosen depending upon the
size of the network such that the random variable lies between $0$ and $1$.
The same initial condition is used for both the layers in
the multiplex network. A very narrow width of the function leads to
almost very close value of the initial conditions for a fraction of nodes 
leading to the coherent state.
 
  \begin{figure}[t]
  \centerline{\includegraphics[width=3.5in, height=2.0in]{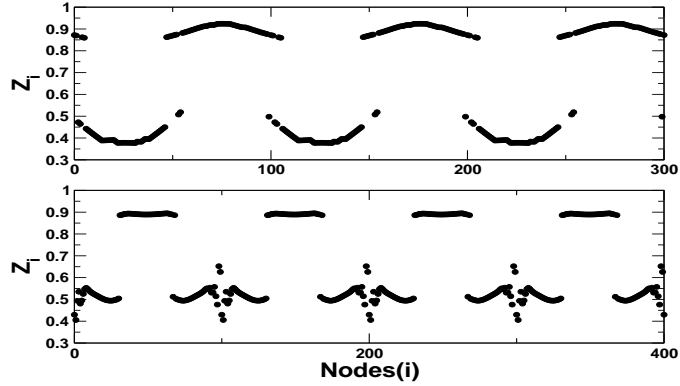}}
  \caption{Snapshots for (a) Three layer and (b) Four layer multiplex network. Parameters are $\varepsilon = 0.28$ and $r=0.32$. Number of nodes $N=100$ for each layer.}
  \label{fig.mul_lay}
  \end{figure}

In the absence of any coupling between the nodes ($\varepsilon=0$) or for 
weak couplings, all the nodes evolve independently and no spatial coherence is observed. For instance, as demonstrated in Fig.~\ref{fig.main}(a), for $\varepsilon=0.1$ , the evolution of the nodes in the multiplex network yields an 
incoherent state with no correlations in the neighboring nodes. 
As the coupling strength is increased, a partially coherent state emerges at 
$\varepsilon=0.28$ with correlated spatial values of the neighboring nodes in 
the end and in the middle regions of each layer, however, the spatial
range of the incoherent region is more than the coherent region (Fig.~\ref{fig.main}(b)). This coexistence of the coherent and incoherent dynamics corresponds to the chimera state in the multiplex network.

\begin{figure}[t]
\centering
\includegraphics[width=3.5in, height=2.0in]{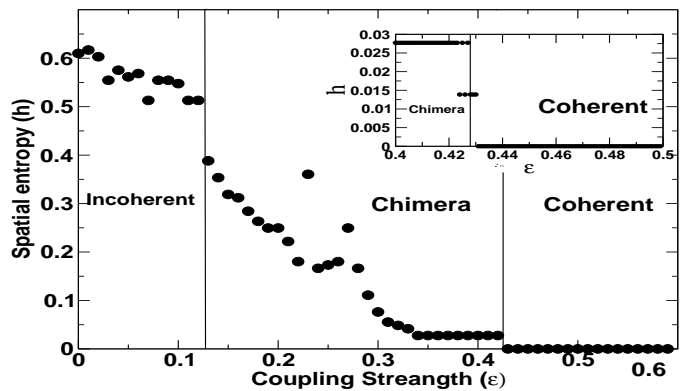}
\caption{Spatial entropy as a function of $\varepsilon$. Figure are plotted for $N=100$ and $r=0.32$ }
\label{fig.trans}
\end{figure}

The dynamical behaviour of two layers of a multiplex network is a replica of each other manifesting exactly the same spatio-temporal patterns (Fig.~\ref{fig.main}). Exactly same behaviour is observed for multiplex networks having more than two layers (Fig.\ref{fig.mul_lay}).

 At the same coupling value, the spatio-temporal dynamics (Fig.~\ref{fig.main} (g)) reflects non-regular skeletal type pattern in the incoherent regions. This irregularity of the pattern suggests that, in the multiplex network framework, a node may get attracted to either 
of the upper or lower region depending on its initial value as reported for the isolated network \cite{theo9}. 

As we increase the coupling strength further, range of the 
incoherent region decreases as depicted by Fig.~\ref{fig.main}(c) for $\varepsilon=0.3$. At $\varepsilon=0.4$, we observe a sharp discontinuity in the 
otherwise smooth profile of $z(j)$ and the incoherency appears at two distinct points in each layer. This is a bifurcation point for the coherent-incoherent transition. Above this coupling value, all the nodes in the multiplex network acquire the complete coherent state as indicated by the appearance of a smooth geometric profile at $\varepsilon=0.44$ (Fig.~\ref{fig.main}(e)). Fig.\ref{fig.main} depicts spatial regions of incoherent nodes and thus indicates a non-zero spatial entropy with the periodic temporal dynamics, representing a Type II chimera state. Further, the regions of incoherence in the spatial profile continues to exist for narrower intervals with an increase in the coupling strength (Fig.~\ref{fig.main}).

Furthermore, in the Chimera state, the time evolution of all the nodes in the network depict periodic behaviour with the periodicity two depicting
temporal regularity. The coupled dynamics displays the spatial chaos which is defined by the non-zero spatial entropy given by $h=d$ $\log_{e}(2)$, where $d$ represents fraction of the incoherent nodes in network \cite{theo9,spt_entrpy}. 
We show that the distance measure (Eq.\ref{eq.dmsr}) is able to easily distinguish between coherent and chimera state. The discontinuous spatial profile (Fig.~\ref{fig.dismsr1}(b)) at $\varepsilon=0.4$ gives rise to the kinks (Fig.~\ref{fig.dismsr1}(d)) in the distance measure distribution signifying transition to the chimera state. We calculate the spatial entropy as a function of the coupling strength ($\varepsilon$) in order to demonstrate the transition between the chimera to the coherent state. A transition from the chimera to coherent state is indicated by the discontinuous change in spatial entropy of the network (Fig.\ref{fig.trans}).

\section{Multiplex network versus isolated network}
\begin{figure}[t]
\centerline{\includegraphics[width=3.6in, height=2in]{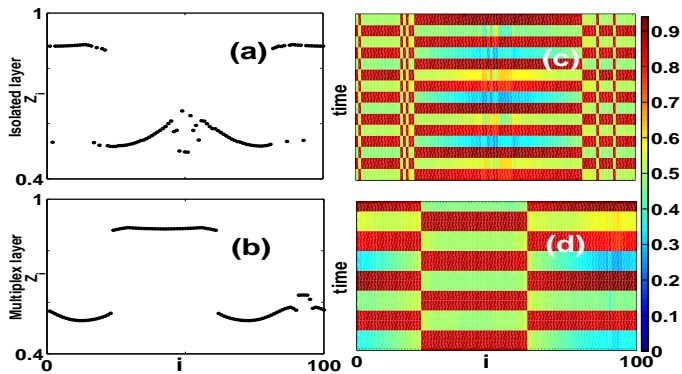}}
\caption{Snapshot and Spatio temporal Pattern for (a,c) Single and (b,d) first $S^1$ ring of multiplexed network are shown. Parameters are $\varepsilon = 0.30$ and $r=0.32$. Number of nodes $N=100$ for each layer. Initial transient is taken as 5000. results are presented for time range 5000 to 5015.}
\label{fig:mul_effect}
\end{figure}
 In order to see the impact of multiplexcity on the dynamical behavior of 
nodes, we compare the dynamical state of an isolated 1D lattice with that of 
the multiplexed with another 1D lattice. We find that while the chimera state is retained after multiplexing, the dynamical evolution of the network differs. 
The multiplexing may enhance or suppress the incoherency. For instance, 
at $\varepsilon=0.30$, the isolated network displays the chimera state with 
incoherence in the middle regions of the spatio temporal pattern 
(Fig.~\ref{fig:mul_effect} (a) and (c)).
After multiplexing, the region of incoherence shrinks to a point 
discontinuity in the middle and intermediate regions 
(Fig.~\ref{fig:mul_effect}). Thus, multiplexing here retains the 
chimera state as well as the type of the chimera state, but leads to 
a change in the 
region of incoherence as well as in the dynamical evolution. 

\section{Sensitivity to initial conditions}  
  \begin{figure}[t]
  \centerline{\includegraphics[width=3.2in, height=2.0in]{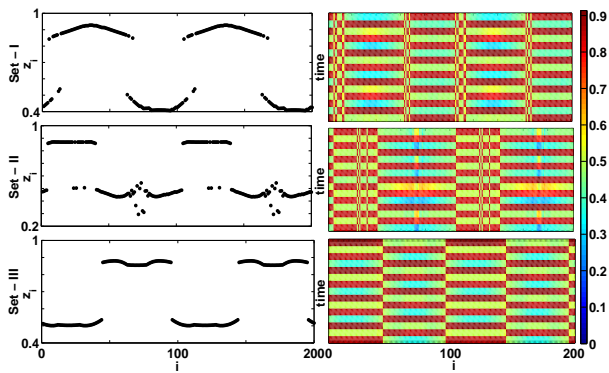}}
  \caption{Spatio temporal pattern for multiplex $S^1$ rings for different 
realizations of the initial conditions. Parameters are $\varepsilon = 0.32$ and $r=0.32$. Number of nodes $N=100$ for each layer. Initial transient is taken as 5000. results are presented for time range 5000 to 5015.}
  \label{fig:initial}
  \end{figure}

Furthermore, similar to the isolated networks, in the multiplex networks as 
well, the chimera exhibits dependency on the initial conditions. 
As already discussed in the section III, the 
spatial profile displayed by the chimera state can be very different for
different profile of initial conditions for the same set of control parameters, namely 
$\mu$ and $r$. But interestingly, even for the initial condition given by the 
same profile as Eq.~\ref{ini_eq} with a constant value $\sigma$ for a given
network size $N$, different realizations of the initial conditions
can lead to a different incoherent regions. 
For example, at $\varepsilon=0.30$, for three different realizations of the 
initial conditions, all given by Eq.~\ref{ini_eq}, three different 
spatio temporal patterns are observed (Fig. \ref{fig:initial}). Though, the multi-chimera state is evident for all the realizations, the region of incoherence differs without any consistent behavior.

\begin{figure}[t]
\centerline{\includegraphics[width=3.3in, height=2.0in]{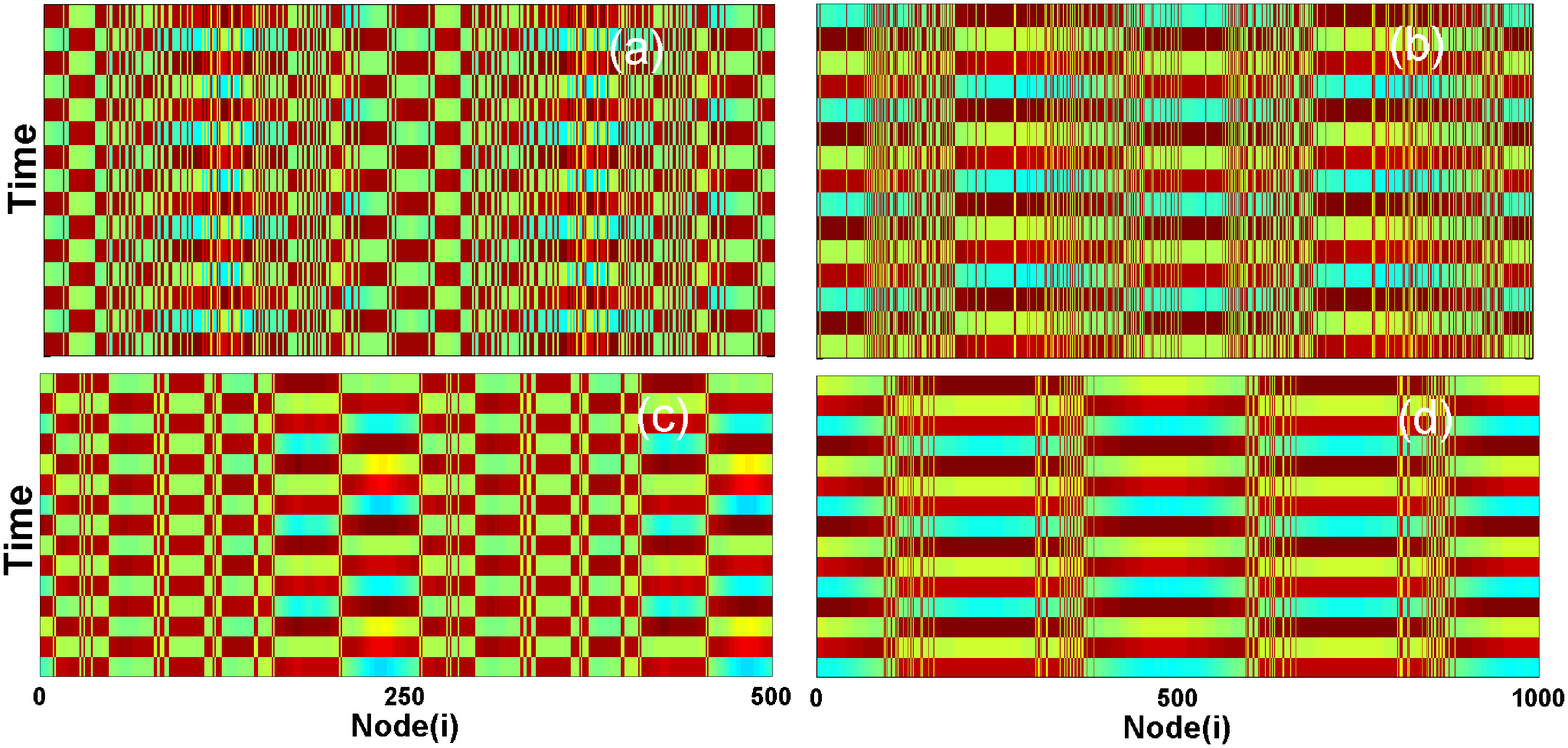}}
\caption{Spatio temporal pattern for multiplex $S^1$ rings with coupling strength (a) $\varepsilon=0.21$ (b) $\varepsilon=0.23$ (c) $\varepsilon=0.28$ and
(d) $\varepsilon=0.27$ . Coupling range $r=0.32$ and network size for (a),(c) 
$N=500$ and (b),(d) $N=1000$. Initial transient is taken as 5000. results are presented for time range 5000 to 5015.}
\label{fig:n_size}
\end{figure}
\begin{figure}[t]
\centerline{\includegraphics[width=3.2in, height=2.0in]{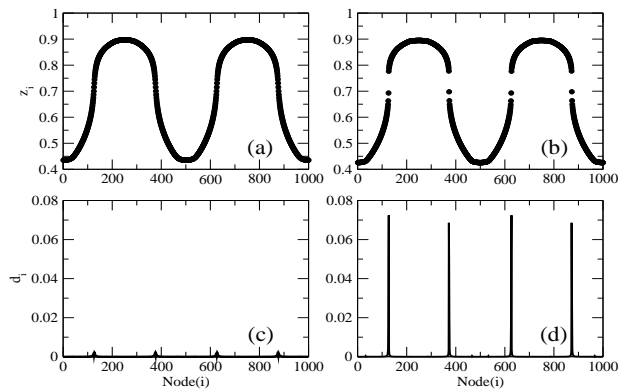}}
\caption{Distance measure for multiplex $S^1$ ring network.  (a),(c) for
$\varepsilon=0.44$ and (b),(d) for $\varepsilon=0.41$. 
Network size $N=500$ and coupling range $r=0.32$ in each layer.}
\label{fig:n_sizemsr}
\end{figure}

Moreover, we investigated the impact of network size on 
emergence of chimera state as well as on the critical coupling strength below
which one observes incoherent-coherent regime.
Fig. \ref{fig:n_size} presents chimera state for a multiplex ring network with 
two different network size. For both the network sizes, the coupled evolution
 exhibits the coexistence of coherence-incoherence dynamics. 
However, the critical coupling strength for 
the coherent-incoherent bifurcation increases to 
$\varepsilon =0.41$ (Fig.\ref{fig:n_sizemsr}) as compared to $\varepsilon = 0.4$ for network size $N=200$ as indicated by Fig.~\ref{fig.main}. 

\section{Conclusion}
To summarize, we report an emergence of the chimera in the multiplex networks 
with the layers being represented by 1-d lattice architecture having non-local 
couplings. We find that an emergence of the chimera is identical in the mirror 
layers arising due to the underlying symmetry of the network. Furthermore, 
while the temporal behavior of the network remains periodic even after 
multiplexing, the range of the coupling strength for which chimera is 
observed changes.  The chimera in the multiplex network is found to be 
sensitive to the changes in the initial conditions, which is revealed through 
the changes in the incoherent region of the dynamical evolution for 
different sets of initial conditions. We also show that 
the critical coupling strength increases with the size of the network. 
The results presented here may provide a better understanding to the 
peculiar nature of the chimera state observed in many natural systems like unihemispheric sleep, ventricular fibrillation, brain networks which incorporates multilayer network architecture \cite{2D_4}.
\section{Acknowledgments} 
SJ is grateful to Department of Science and Technology (DST), Government of India and Council of Scientific and Industrial Research, Government of India project grants
EMR/2014/000368 and 25(0205)/12/EMR-II respectively for financial support. SG acknowledges DST, Government of India, for the INSPIRE fellowship (IF150149) as well as the Complex Systems Lab members for timely help and useful discussions. Authors acknowledge Conference on Nonlinear Systems and Dynamics 2015 held at IISER Mohali where this work was initiated.


\end{document}